\pdfoutput=1
 \documentclass[aps,prl,twocolumn,showpacs,superscriptaddress,preprintnumbers,longbibliography]{revtex4-1}  
\usepackage[normalem]{ulem}
\usepackage{amsmath}
\usepackage{amssymb}
\usepackage{epsfig}
\usepackage{graphicx}
\usepackage{hyperref}
\usepackage{tabu}
\usepackage{boldline}
\usepackage{xspace}
\usepackage{slashed}
\usepackage{multirow}
\usepackage{booktabs}
\usepackage{diagbox}
\usepackage{tabularx}
\usepackage{comment}
\usepackage{floatrow}
\newfloatcommand{capbtabbox}{table}[][\FBwidth]
\usepackage{color}
\usepackage[normal]{subfigure}
\usepackage[table]{xcolor}
\usepackage{enumitem}
\usepackage[utf8]{inputenc}
\usepackage{colortbl}
\definecolor{nicered}{rgb}{0.7,0.1,0.1}
\definecolor{nicegreen}{rgb}{0.1,0.5,0.1}
\definecolor{red}{rgb}{1.0, 0, 0}
\definecolor{pink}{RGB}{255, 0, 145}

\definecolor{LightCyan}{rgb}{0.88,1,1}
\definecolor{piggypink}{rgb}{0.99, 0.87, 0.9}
\definecolor{applegreen}{rgb}{0.55, 0.71, 0.0}
\definecolor{darkpastelgreen}{rgb}{0.01, 0.75, 0.24}
\definecolor{green-yellow}{rgb}{0.68, 1.0, 0.18}

\newcommand{\beq}{\begin{equation}}
\newcommand{\eeq}{\end{equation}}
\newcommand{\beqa}{\begin{eqnarray}}
\newcommand{\eeqa}{\end{eqnarray}}

\begin{document}

\title{SENSEI: Characterization of Single-Electron Events Using a Skipper-CCD
}

\author{The SENSEI Collaboration: \\ Liron Barak}

\affiliation{\normalsize\it 
 School of Physics and Astronomy, 
 Tel-Aviv University, Tel-Aviv 69978, Israel}

 \author{Itay M. Bloch}
 \affiliation{\normalsize\it 
 School of Physics and Astronomy, 
 Tel-Aviv University, Tel-Aviv 69978, Israel}

\author{Ana Botti}
\affiliation{\normalsize\it 
Department of Physics, FCEN, University of Buenos Aires and IFIBA, CONICET, Buenos Aires, Argentina}
\affiliation{\normalsize\it 
Fermi National Accelerator Laboratory, PO Box 500, Batavia IL, 60510, USA}

\author{Mariano Cababie}
\email{mcababie@df.uba.ar}
\affiliation{\normalsize\it 
Department of Physics, FCEN, University of Buenos Aires and IFIBA, CONICET, Buenos Aires, Argentina}
\affiliation{\normalsize\it 
Fermi National Accelerator Laboratory, PO Box 500, Batavia IL, 60510, USA}

\author{Gustavo Cancelo}
\affiliation{\normalsize\it 
Fermi National Accelerator Laboratory, PO Box 500, Batavia IL, 60510, USA}

\author{Luke Chaplinsky}
\affiliation{\normalsize\it 
C.N.~Yang Institute for Theoretical Physics, Stony Brook University, Stony Brook, NY 11794, USA}
\affiliation{\normalsize\it 
Department of Physics and Astronomy, Stony Brook University, Stony Brook, NY 11794, USA}

\author{Fernando Chierchie}
\affiliation{\normalsize\it 
Fermi National Accelerator Laboratory, PO Box 500, Batavia IL, 60510, USA}

\author{Michael Crisler}
\affiliation{\normalsize\it 
Fermi National Accelerator Laboratory, PO Box 500, Batavia IL, 60510, USA}

\author{Alex Drlica-Wagner}
\affiliation{\normalsize\it 
Fermi National Accelerator Laboratory, PO Box 500, Batavia IL, 60510, USA}
\affiliation{\normalsize\it Kavli Institute for Cosmological Physics, University of Chicago, Chicago, IL 60637, USA}
\affiliation{\normalsize\it  Department of Astronomy and Astrophysics, University of Chicago, Chicago IL 60637, USA}

 \author{Rouven Essig}
\affiliation{\normalsize\it 
C.N.~Yang Institute for Theoretical Physics, Stony Brook University, Stony Brook, NY 11794, USA}

 \author{Juan Estrada}
\affiliation{\normalsize\it 
Fermi National Accelerator Laboratory, PO Box 500, Batavia IL, 60510, USA}

\author{Erez Etzion}
\affiliation{\normalsize\it 
 School of Physics and Astronomy, 
 Tel-Aviv University, Tel-Aviv 69978, Israel}

\author{Guillermo Fernandez Moroni}
\affiliation{\normalsize\it 
Fermi National Accelerator Laboratory, PO Box 500, Batavia IL, 60510, USA}

\author{Daniel Gift}
\affiliation{\normalsize\it 
C.N.~Yang Institute for Theoretical Physics, Stony Brook University, Stony Brook, NY 11794, USA}
\affiliation{\normalsize\it 
Department of Physics and Astronomy, Stony Brook University, Stony Brook, NY 11794, USA} 

\author{Stephen E. Holland}
\affiliation{\normalsize\it 
Lawrence Berkeley National Laboratory, One Cyclotron Road, Berkeley, California 94720, USA}

\author{Sravan Munagavalasa}
\affiliation{\normalsize\it 
C.N.~Yang Institute for Theoretical Physics, Stony Brook University, Stony Brook, NY 11794, USA}
\affiliation{\normalsize\it 
Department of Physics and Astronomy, Stony Brook University, Stony Brook, NY 11794, USA}

 \author{Aviv Orly}
\affiliation{\normalsize\it 
 School of Physics and Astronomy, 
 Tel-Aviv University, Tel-Aviv 69978, Israel}

\author{Dario Rodrigues}
\affiliation{\normalsize\it 
Department of Physics, FCEN, University of Buenos Aires and IFIBA, CONICET, Buenos Aires, Argentina}
\affiliation{\normalsize\it 
Fermi National Accelerator Laboratory, PO Box 500, Batavia IL, 60510, USA}

\author{Aman Singal}
\affiliation{\normalsize\it 
Department of Physics and Astronomy, Stony Brook University, Stony Brook, NY 11794, USA}

 \author{Miguel Sofo Haro}
\affiliation{\normalsize\it 
Fermi National Accelerator Laboratory, PO Box 500, Batavia IL, 60510, USA}
\affiliation{Centro At\'omico Bariloche, CNEA/CONICET/IB, Bariloche, Argentina}

\author{Leandro Stefanazzi}
\affiliation{\normalsize\it 
Fermi National Accelerator Laboratory, PO Box 500, Batavia IL, 60510, USA}

\author{Javier Tiffenberg}
\affiliation{\normalsize\it 
Fermi National Accelerator Laboratory, PO Box 500, Batavia IL, 60510, USA}

\author{Sho Uemura}
\affiliation{\normalsize\it 
 School of Physics and Astronomy, 
 Tel-Aviv University, Tel-Aviv 69978, Israel}

\author{Tomer Volansky}
\affiliation{\normalsize\it 
 School of Physics and Astronomy,   
 Tel-Aviv University, Tel-Aviv 69978, Israel}

\author{Tien-Tien Yu}
\affiliation{\normalsize\it 
Department of Physics and Institute for Fundamental Science, University of Oregon, Eugene, Oregon 97403, USA}

\date{\today}

\preprint{FERMILAB-PUB-21-278-E}

\begin{abstract}
\noindent 
We use a science-grade Skipper Charge Coupled Device (Skipper-CCD) operating in a low-radiation background environment to develop a semi-empirical model that characterizes the origin of single-electron events in CCDs.
We identify, separate, and quantify three independent contributions to the single-electron events, which were previously bundled together and classified as ``dark counts'': dark current, amplifier light, and spurious charge. 
We measure a dark current, which depends on exposure, of $(5.89 \pm 0.77) \times 10^{-4} \ e^{-}/{\rm pix}/{\rm day}$, and an unprecedentedly low spurious charge contribution of $(1.52 \pm 0.07) \times 10^{-4} \ e^{-}/{\rm pix}$, which is exposure-independent. 
In addition, we provide a technique to study events produced by light emitted from the amplifier, which allows the detector's operation to be optimized to minimize this effect to a level below the dark-current contribution. 
Our accurate characterization of the single-electron events allows one to greatly extend the sensitivity of experiments searching for dark matter or coherent neutrino scattering. 
Moreover, an accurate understanding of the origin of single-electron events is critical to further progress in ongoing R\&D efforts of Skipper and conventional CCDs. 
\end{abstract}

\maketitle

\section{\label{sec:introduction} INTRODUCTION} 

Charged coupled devices (CCDs) are widely used photon detectors for scientific purposes since their invention in 1969~\cite{Smith2010,janesick}. 
More recently, they have also been adopted as particle detectors in rare-event searches for coherent neutrino scattering~\cite{CONNIE2} and dark matter particles~\cite{Barreto:2011zu,DAMIC2019}.
The development of the Skipper-CCD~\cite{javier2017}, with its deep sub-electron read-out noise and resulting ability to detect events producing only a single photon or a single electron, has revolutionized the search for coherent neutrino scattering and dark matter particles, since these produce events typically containing only one or a few ionized electrons~\cite{Freedman:1973yd,RouvenTomerER}. 
The first science results from data collected with a single Skipper-CCD have already been presented in~\cite{SENSEI2018,SENSEI2019,SENSEI2020}, and the construction of large-mass detectors for both neutrino and dark matter particles searches is underway or planned~\cite{fernandez2020,SENSEI2020,DAMIC-M2020,OSCURA}.

The data from~\cite{SENSEI2018,SENSEI2019,SENSEI2020} contains a large number of single-electron events (SEE).  A detailed understanding of the nature and origin of these SEE in silicon crystals, and in (Skipper-)CCDs in particular, is crucial both for maximizing the sensitivity to coherent neutrino scattering and to dark matter and for assessing the discovery potential of these experiments. 

We present in this paper a set of techniques to identify and characterize the nature and origin of SEE utilizing a science-grade Skipper-CCD.  We employ a Skipper-CCD from the first batch of sensors fabricated using ultra high-resistivity silicon ($R>18$~k$\Omega$). We also propose a semi-empirical model that describes the observed SEE.
The methodology described here provides a set of techniques that we expect can be implemented with any Skipper-CCDs. 
Furthermore, it enables current and planned experiments utilizing Skipper-CCDs, such as CONNIE, SENSEI, DAMIC-M, and Oscura, to mitigate these background events and greatly enhance their sensitivity to, and discovery potential of, coherent neutrino scattering and sub-GeV dark matter. 
\vspace{-3mm}

\section{\label{sec:expsetup} Technical description}

\vspace{-3mm}
We provide here a brief description of the Skipper-CCD and the data taking.

\subsection{\label{sub:skp} The Skipper-CCD}

The measurements presented in this work were performed using a science-grade Skipper-CCD from the SENSEI experiment~\cite{SENSEI2020}. The sensors were designed by the Lawrence Berkeley National Laboratory (LBNL), and fabricated at Teledyne/DALSA using high-resistivity ($>$18k$\Omega$-cm) silicon wafers with a thickness of 675~$\mu$m. The packaging and testing of the sensors was done at the SiDet facility at the Fermi National Accelerator Laboratory (FNAL) using the package design described in~\cite{SENSEI2020}. 
The sensor was installed in a vacuum vessel that was placed in a low-radiation environment at the MINOS underground facility at FNAL. The shielding surrounding the Skipper-CCD can be divided into an ``internal'' and an ``external'' shield, corresponding to whether it was installed inside or outside of the vacuum vessel that houses the Skipper-CCD. An external shield was added in the midst of the data-taking process.
For details on both types of shielding see~\cite{SENSEI2020}. 

\renewcommand{\arraystretch}{1.2}
\begin{table}[b]
\centering
\begin{tabular}{lcc}
\hline
                          & Value                  & Units                 \\ \hline
CCD dimensions            & 6144  $\times$  886    & pixels              \\
Pixel Size                & 15 $ \times$ 15        & $\mu$m$^{2}$                \\
Thickness                 & 675                    & $\mu$m                \\
Total mass                & 1.926                  & g                \\
CCD temperature           & 135                    & K              \\
Number of amplifiers      & 4 (2 used)          &                       \\
Readout time (1 sample)   & 42.825                 & $\mu$s                \\
Readout noise (1 sample)  & 2.5                    & $e^{-} \ {\rm rms} \ / \ {\rm pix} $ \\
Readout noise (n samples) & $\frac{2.5}{\sqrt{n}}$ & $e^{-} \ {\rm rms} \ / \ {\rm pix}$  \\ \hline
\end{tabular}
\caption{Main characteristics of the Skipper-CCD used in this work. \label{tab:sensor}}
\end{table}

The Skipper-CCD used for this work has four identical readout stages, one in each corner.  The usual mode of operation is to read out one quarter (a quadrant) of the Skipper-CCD through the corresponding sensor, simultaneously and independently of the readout of the other quadrants.  However, there are no physical barriers between the quadrants and it is possible to read the entire Skipper-CCD though any one of the readout stages if desired. The active area is a standard LBNL three-phase CCD with a buried $p$-channel fabricated on a high-resistivity $n$-type bulk \cite{holland2009}. 
Each column of pixels is separated by ``channel stops'', highly doped regions that prevent the spread of the charges from one column into another.
The fabrication process for this detector was optimized for dark matter searches, and no polishing of the backside (usually referred to as ``thinning'') was done to preserve as much target mass as possible. The silicon bulk is fully depleted at approximately 40~V but was operated at a 70~V substrate voltage to reduce the diffusion of charge to neighboring pixels in the bulk. 

The detector was operated at low temperature (135~K) to reduce the probability of surface and bulk electrons to be promoted into the conduction band due to thermal agitation~\cite{janesick}.
At this temperature, the fraction of thermally-induced events, which we refer to as ``intrinsic dark current'', is sub-dominant compared to other sources of charge~\cite{SENSEI2020}. 
Operating the sensor at even lower temperatures had the undesirable effect of increasing the charge transfer inefficiency. 
As there was an unusual excess of SEE in the third quadrant and a high charge transfer inefficiency in the fourth quadrant, the data collected by these two quadrants was not used in the final analysis. 
A brief summary of the Skipper-CCD's characteristics is given in Table~\ref{tab:sensor}. 

\begin{figure}[t]
  \includegraphics[width=.9\linewidth]{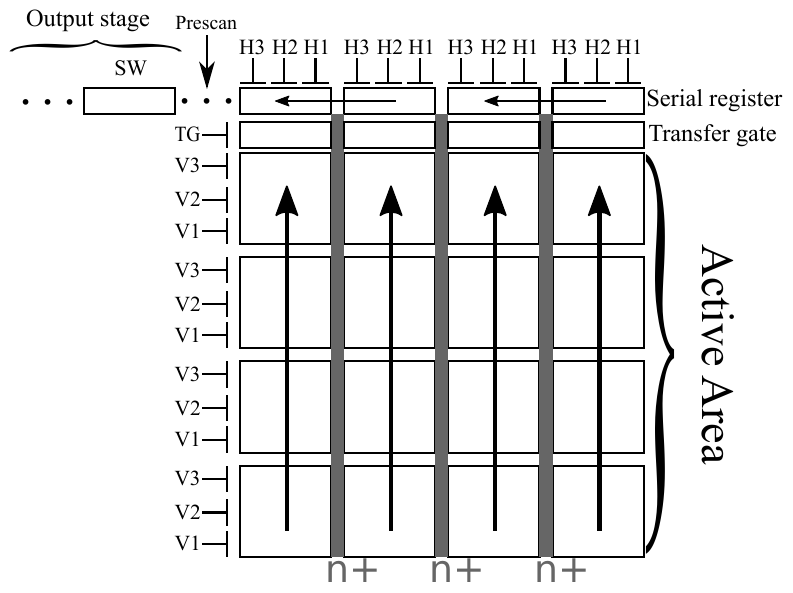}
  \caption{Schematic illustration of a $4\times4$ pixels quadrant of a CCD. The arrows show the direction in which collected charges are transferred during readout. A detailed schematic of the readout stage is provided in Fig.~\ref{fig:RO}. H1, H2 and H3 are the last horizontal clocks in the serial register before the Summing Well (SW). In gray, highly doped $n$-type ``channel stops''.
  \label{fig:tech.1}}
\end{figure}

\begin{figure}[b]
  \includegraphics[width=1\linewidth]{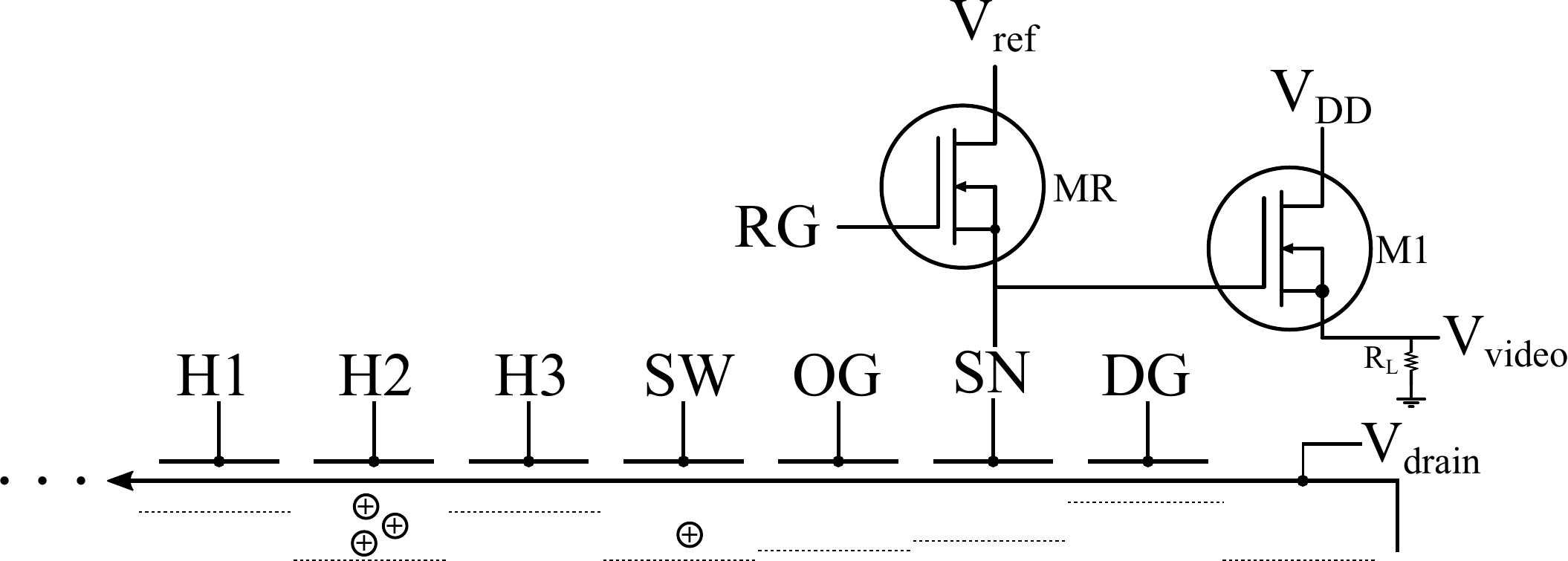}
  \caption{Schematic illustration of a Skipper-CCD readout stage \cite{Tiffenberg2017}. H1, H2 and H3 are the last horizontal clocks in the serial register before the Summing Well (SW). }
  
  \label{fig:RO}
\end{figure}

As shown in Fig.~\ref{fig:tech.1}, each quadrant of the CCD can be divided into four regions: active area, transfer gate, serial register, and readout stage. 
The active area of each quadrant is an array of $3072\times 443$ pixels where the charges are collected during the exposure of the CCD sensor. 
A schematic depiction of the readout stage is provided in Fig.~\ref{fig:RO}.

\subsection{\label{sub:datataking} Data-taking cycle}

The standard data-taking cycle of a CCD consists of three phases:

\begin{itemize}
\item {\bf Cleaning.}  During this phase, the substrate bias of the CCD is set at $0$~V \cite{holland2009} whereas voltages from the vertical clocks on the surface of the CCD are set at $9$~V in order to invert the surface with electrons from the channel stops \cite{holland2003}, filling the surface traps and thereby reducing intrinsic dark current and “resetting” the  CCD. At the end of this procedure, the substrate voltage is set back to the depletion voltage and the surface clocks are set to their previous configuration. The entire CCD is then read out in order to remove all of the accumulated charge so that one can start the next phase with a ``clean" CCD.
\item {\bf Exposure.} During this phase all voltages are kept fixed. Any ionizing interaction that occurs within the active area of the detector therefore leaves electrons trapped inside the pixels, which are later read during the readout phase. 
V$_{\rm DD}$, the bias voltage for the output transistor M1, is set to 0 during this phase and turned on during the other phases.
\item {\bf Readout.}  The collected charges are transferred vertically row by row into the serial register through the transfer gate. Once the charges reach the serial register, they are transported horizontally, one pixel at a time, to the readout stage (Fig.~\ref{fig:tech.1}). At the readout stage, the charges collected in each pixel are converted into a voltage signal that is finally measured by the readout electronics (Fig.~\ref{fig:RO}).
\end{itemize}

\section{\label{sec:empmodel} Semi-empirical model for the origin of the single-electron events}

The SEE per pixel have two main contributions that scale with time: events produced during Exposure ($\mu_{\rm EXP}$) and events produced during Readout ($\mu_{\rm RO}$). 
The rates of these contributions 
are expected to be different as the amount of light produced by the readout stage depends on the base current of the output transistor, which varies during Readout and is turned off during the Exposure phase under normal operating  conditions.
We identify also a time-independent contribution, which we define as spurious charge ($\mu_{\rm SC}$). Spurious charge is generated when the surface voltages of the pixels in the active area, transfer gate, and serial register change in order to transfer charge from one pixel to the other.
Such events are produced during the Cleaning and Readout phases and will be discussed later on. 
Since they depend solely on the number of times a pixel is clocked (\emph{i.e.}, the number of times a pixel transfers charges to its neighbor), they are modeled as an additional exposure-independent term.

Following these definitions, the total number ($\mu$) of SEE per pixel generated during a data-taking cycle can be written as
\begin{equation}
\begin{split}
\label{eq1.001}
\mu(t_{\rm EXP},t_{\rm RO}) & =\mu_{\rm EXP} (t_{\rm EXP})+\mu_{\rm RO}(t_{\rm RO})+\mu_{\rm SC} \\
                             & = \lambda_{\rm EXP} t_{\rm EXP}+ \lambda_{\rm RO} t_{\rm RO} +  \mu_{\rm SC}\,,
\end{split}
\end{equation}
where in the second line we 
assumed that both $\mu_{\rm EXP}$ and $\mu_{RO}$ scale linearly with time. 
The parameters $\lambda_{\rm EXP}$  and $\lambda_{\rm RO}$ are the SEE rates 
during exposure and readout phase, respectively, in units of events per pixel per day, while 
the times $t_{\rm EXP}$ and $t_{\rm RO}$ are expressed in days. 

There are three contributions to the SEE: dark current, amplifier light, and spurious charge. 
In the following subsection we discuss each of these contributions in detail. 
In that discussion, these contributions will be classified based on the following primary characteristics: spatial distribution (localized or uniform) and time-dependence.
Throughout our analysis we use the event-selection criteria discussed in~\cite{SENSEI2020} and summarized in Table~I of that paper,
with the exception of the edge mask and the halo mask (cuts number 6 and 9 of Table~I of~\cite{SENSEI2020}), which we set to 40 pixels (instead of 60) to increase the statistics for this work.

\subsection{\label{sub:DC} Dark current}

Although the CCD is inside a sealed vacuum vessel, SEE can be generated in the CCD during both Exposure and Readout, even in the absence of any external source (except unavoidable environmental radiation interactions) and is unrelated to charges being shifted from pixel-to-pixel. We refer to SEE generated in this way as dark current (DC) and 
differentiate between two distinct contributions: intrinsic and extrinsic, depending on whether the current is generated by the CCD itself (intrinsic) or through an interaction with its environment (extrinsic). 
DC contributes during both the Exposure and Readout phases of data-taking.
The number of events coming from DC scales linearly with time during the Exposure phase and hence $\lambda_{\rm DC}$, the rate of SEE generated from DC , will be one of the components of $\lambda_{\rm EXP}$.
During the Readout phase, the exposure of the pixels is non-uniform as the last pixel read has an additional exposure time of $t_{\rm RO}$ when compared to the pixel that is read first. 
Overall, the average contribution of the DC during the  Readout phase is given by $\lambda_{{\rm DC}}/2$.

\subsubsection{\label{sub:iDC} Intrinsic dark current}

Intrinsic dark current (usually referred plainly as dark current in the literature) is the most well-studied current contribution for SEE ~\cite{widenhorn2002,janesick}. 
This source generates, via thermal fluctuations across the silicon band-gap, single electron-hole pairs. 
The holes, which we collect using a $p$-buried channel, are trapped in the potential wells within the pixels. Since this thermal agitation is a random process independent of time and space, the number of 
holes collected from the thermal agitation in a pixel for a given period of time is a random variable that follows a Poisson distribution. 
The expected value is the intrinsic DC rate multiplied by the exposure time of the pixel. 
We can further differentiate between two types of intrinsic DC that appear in a buried-channel CCD: bulk and surface intrinsic DC.
The latter component can be greatly reduced, at least temporarily, with an inversion of the surface in the Cleaning phase~\cite{holland2003}. This empties the traps that mediate the generation of surface DC. The traps begin to fill again during the subsequent Exposure phase, but this recovery is inhibited at low temperature.
Estimates from the model developed in~\cite{burke1991} show that less than 0.1$\%$ of the surface intrinsic DC is recovered at the operating temperature of 135~K after 24 hours (much longer than any measurement carried out during this work).
Therefore, we find that the bulk component dominates the intrinsic DC.

\subsubsection{\label{sub:EC} Extrinsic dark current}

In~\cite{SENSEI2020}, we reported a SEE rate of $(1.594 \ \pm \ 0.160) \times 10^{-4} \ e^{-}/{\rm pix}/{\rm day}$, which, despite being the lowest value ever reported in a CCD, is at least one order of magnitude above the theoretically expected intrinsic DC rate at 135~K. 
This discrepancy is explained by additional environmentally-induced contributions to the observed SEE, distributed approximately uniformly across the CCD (after applying the event-selection criteria and with the given statistical uncertainties) and, as the intrinsic DC, increasing linearly with the exposure time.
In this sense it is, in principle, indistinguishable from the intrinsic DC. 
This extrinsic contribution was identified for priorly in~\cite{SENSEI2020} and is directly related to the environmental radiation. A detailed discussion of the physical processes that likely explain these SEE is discussed in~\cite{du2020sources}.

\subsection{\label{sub:ampInduced} SEE produced by amplifier light}

As the number of SEE per pixel decreases with lower temperature and background radiation as well as improved detector performance, additional low-energy signals appear that are not contained in the definition of DC. 
This collaboration recently reported an excess of SEE near the readout stage~\cite{SENSEI2019}, which we refer to as amplifier-light events.
This effect was previously investigated by many authors~\cite{bartelink1963,toriumi1987,lanzoni1991,bude1992,tsang1997} and is due to infrared photons emitted by the M1 transistor of the readout stage (see Fig \ref{fig:RO}). 
Because the photons have a finite range in silicon and are emitted continuously, charge generated by this effect is spatially localized in the region near the readout stage and increases linearly with time.
We express the corresponding rate as $\lambda_{\rm AL}$ (in units of events per pixel per day)
As this contribution is localized, $\lambda_{\rm AL}$ depends not only on the distance from the readout stage but on which zone of the CCD is under study. 
For the sake of simplicity, we average $\lambda_{\rm AL}$ over the whole CCD and do not study the spatial dependence.
  
Since V$_{\rm DD}$ is set to 0 during Exposure, SEE due to amplifier light are only produced in Readout and Cleaning phases. 
Note that if V$_{\rm DD}$ is kept on during the Exposure phase, an additional non-negligible amplifier light contribution must be taken into account.
That contribution may be different, since the voltage in the floating gate (which is also the gate of the M1 transistor, see Fig.~\ref{fig:RO}) is constant during Exposure but changes rapidly during Readout.
Moreover, as the active area is being read and pixels are being transported to the sense node, the spatial dependence of the amplifier light during the Readout and Exposure phases is different.

\subsection{\label{section:SpuriousCharge} Spurious charge }

The last SEE contribution to take into consideration is the spurious charge (SC), which is generated when the voltages in the active area or serial register are clocked~\cite{janesick,haro2016measurement,haro2016taking}. 
As noted earlier, SCs are generated during both Cleaning and Readout phases and depend solely on the number of times a pixel is clocked. 
Therefore, the SC is a spatially uniform, time-independent, intrinsic contribution to the SEE.

\renewcommand{\arraystretch}{1.7}
\begin{table*}[t!]
\centering
\caption{Summary of charge contributions and their properties, following Eq. \ref{eq1.002}. The units for $\lambda_{\rm DC}$ and $\lambda_{\rm AL}$ are $e^{-}/{\rm pix}/{\rm day}$, while $\mu_{\rm SC}$ is in $e^{-}/{\rm pix}$.
At 135~K and after an inversion of the CCD surface, the intrinsic DC is dominated by the bulk intrinsic DC, which has a clear linear time-dependence; we can ignore the surface intrinsic DC which has a non-linear time-dependence. }

\label{table1.2}
\begin{tabular}{|c|c|c|c|c|c|} 
\hline
\multicolumn{2}{|c|}{\multirow{3}{*}{\begin{tabular}[c]{@{}c@{}}\textbf{Contribution}\\\textbf{($e^{-}/{\rm pix}$)}\end{tabular}}}                        & \multicolumn{3}{c|}{\textbf{Time dependence}}                                                                                                          & \multirow{3}{*}{\begin{tabular}[c]{@{}c@{}}\textbf{Spatial}\\\textbf{distribution}\end{tabular}}  \\ 
\cline{3-5}
\multicolumn{2}{|c|}{}                                                              & \multicolumn{2}{c|}{\textbf{\textbf{Linear}}}                                       & \multirow{2}{*}{\textbf{\textbf{\textbf{\textbf{ \ Independent \ }}}}} &                                                                                                   \\ 
\cline{3-4}
\multicolumn{2}{|c|}{}                                                              & \ \ Exposure \ \                  & \ \ Readout \ \                             &                                                                  &                                                                                                   \\ 
\hline
\multirow{2}{*}{\begin{tabular}[c]{@{}c@{}}Dark\\ current\end{tabular}} & Intrinsic & \multirow{2}{*}{$\lambda_{\rm DC} \ t_{\rm EXP}$}  & \multirow{2}{*}{$\frac{\lambda_{\rm DC}}{2} \ t_{\rm RO}$} & \multirow{2}{*}{-}                                               & Uniform                                                                                           \\ 
\cline{2-2}\cline{6-6}
                                                                        & Extrinsic &                                     &                                               &                                                                  & Uniform                                                                                           \\ 
\hline
\multicolumn{2}{|c|}{Amplifier-light current}                                       & -                 & $\lambda_{\rm AL} \ t_{\rm RO}$                     & -                                                                & Localized                                                                                         \\ 
\hline
\multicolumn{2}{|c|}{Spurious charge}                                               & -                                   & -                                             & $\mu_{\rm SC}$                                                   & Uniform                                                                                           \\ 
\hline
\multicolumn{2}{l}{}                                                                & \multicolumn{1}{l}{}                & \multicolumn{1}{l}{}                          & \multicolumn{1}{l}{}                                             & \multicolumn{1}{l}{}                                                                             
\end{tabular}
\end{table*}

\subsection{Empirical model}
Given the three contributions discussed in the previous sections, we can express both $\lambda_{\rm EXP}$ and $\lambda_{\rm RO}$ as follows

\begin{equation}
\label{eq1.002a}
\lambda_{\rm EXP} = \lambda_{\rm DC}\,,
\end{equation}

\begin{equation}
\label{eq1.002b}
\lambda_{\rm RO} = \frac{\lambda_{\rm DC}}{2} +\lambda_{\rm AL}\,,
\end{equation}
and re-write Eq.~\eqref{eq1.001} to obtain our full empirical model:
\begin{eqnarray}
\label{eq1.002}
\mu_{(t_{\rm EXP},t_{\rm RO})}&=& \lambda_{\rm DC} t_{\rm EXP} \nonumber \\& +& \biggl(\frac{\lambda_{\rm DC}}{2} +\lambda_{\rm AL}\biggr) t_{\rm RO} \\ 
&+& \mu_{\rm SC}\,. \nonumber
\end{eqnarray}

A summary of the charge contributions that enter Eq. \eqref{eq1.002} can be found in Table \ref{table1.2}.
In the next section, we use this model to characterize SEE in a SENSEI Skipper-CCD.

\section{\label{sec:methodology} Single-Electron Events Transfer Curves}

Using the model in Eq.~(\ref{eq1.002}), we propose a set of methods to measure $\lambda_{\rm DC}$, $\lambda_{\rm AL}$, and $\mu_{\rm SC}$. These techniques provide the ``transfer curves" for SEE as a function of both exposure and readout time. 
We use two techniques to measure all of the parameters in Eq.~(\ref{eq1.002}):

\begin{enumerate}

\item[I.] {\bf Determination of $\lambda_{\rm DC}$.} We record several images with a range of different exposure times and a fixed readout time. For each image (with a given exposure time), we measure the amount of SEE per pixel.  We then plot the SEE per pixel as a function of the exposure time and perform a linear fit. The slope of this linear function corresponds to $\lambda_{\rm DC}$ as shown in Eq.~(\ref{eq2.mIII}), and the $y$-intercept is the SEE rate from the readout ($\mu_{\rm RO}$ for a fixed readout time $t_{\rm RO}$) plus the SC ($\mu_{\rm SC}$):
\begin{equation}
\label{eq2.mIII}
\mu(t_{\rm EXP})=  \lambda_{\rm DC} \ t_{\rm EXP} + ( \mu_{\rm RO}+  \mu_{\rm SC})\, .
\end{equation}

\item[II.] {\bf Determination of $\lambda_{\rm AL}$ and $\mu_{\rm SC}$.} Using the measured value for $\lambda_{\rm DC}$ obtained by the previous procedure, $\lambda_{\rm AL}$ and $\mu_{\rm SC}$ can be measured by taking multiple images with different readout times and zero exposure time.
To avoid changing the geometry of the active area (and hence the value of $\lambda_{\rm AL}$), $t_{\rm RO}$ is varied by changing the number of samples that are taken per pixel (see Appendix for a study of SC generation in the readout stage). 
For each image (with a given readout time), we measure the amount of SEE per pixel. 
We then plot the SEE as a function of $t_{\rm RO}$ and perform a linear fit. 
The slope of this linear function is $\frac{\lambda_{\rm DC}}{2} +\lambda_{\rm AL}$ as shown in Eq.~\eqref{eq2.mIV}, while the $y$-intercept is $\mu_{\rm SC}$:

\begin{equation}
\label{eq2.mIV}
\mu(t_{\rm RO})=  \biggl(\frac{\lambda_{\rm DC}}{2} +\lambda_{\rm AL} \biggr) t_{\rm RO}  +  \mu_{\rm SC}\, .
\end{equation}
\end{enumerate}

\section{\label{sec:results} RESULTS AND DISCUSSION}

We study four datasets (see Table~\ref{table1.3}):
\begin{itemize}\itemsep -2pt
  \item[\bf{A:}] 
  To determine $\lambda_{\rm DC}$, eight images were taken, each with a different exposure time $t_{\rm EXP}$ but the same readout time $t_{\rm RO}$.
  \item[\bf{B:}] To study the amplifier light contribution, nine datasets were analyzed, each consisting of six zero-exposure images; each set consists of a different ${\rm V_{DD}}$ voltage value applied in the M1 transistor (see Fig.~\ref{fig:RO}), scanning from $-17$~V to $-25$~V in 1~V steps.
  \item[\bf{C\&D:}] To measure $\mu_{\rm SC}$ and $\lambda_{\rm AL}$, seven datasets were used, each consisting of two zero-exposure images and a different readout time $t_{\rm RO}$. We use two different experimental configurations: for the first (dataset C) we set ${\rm V_{DD}}$ at $-22$~V while avoiding the use of an external shield (same as dataset A).  For the second (dataset D) we set the ${\rm V_{DD}}$ at $-21$~V and add an additional external shield (see discussion in the Technical Description section).
\end{itemize}

\begin{figure}
  \includegraphics[width=1\linewidth]{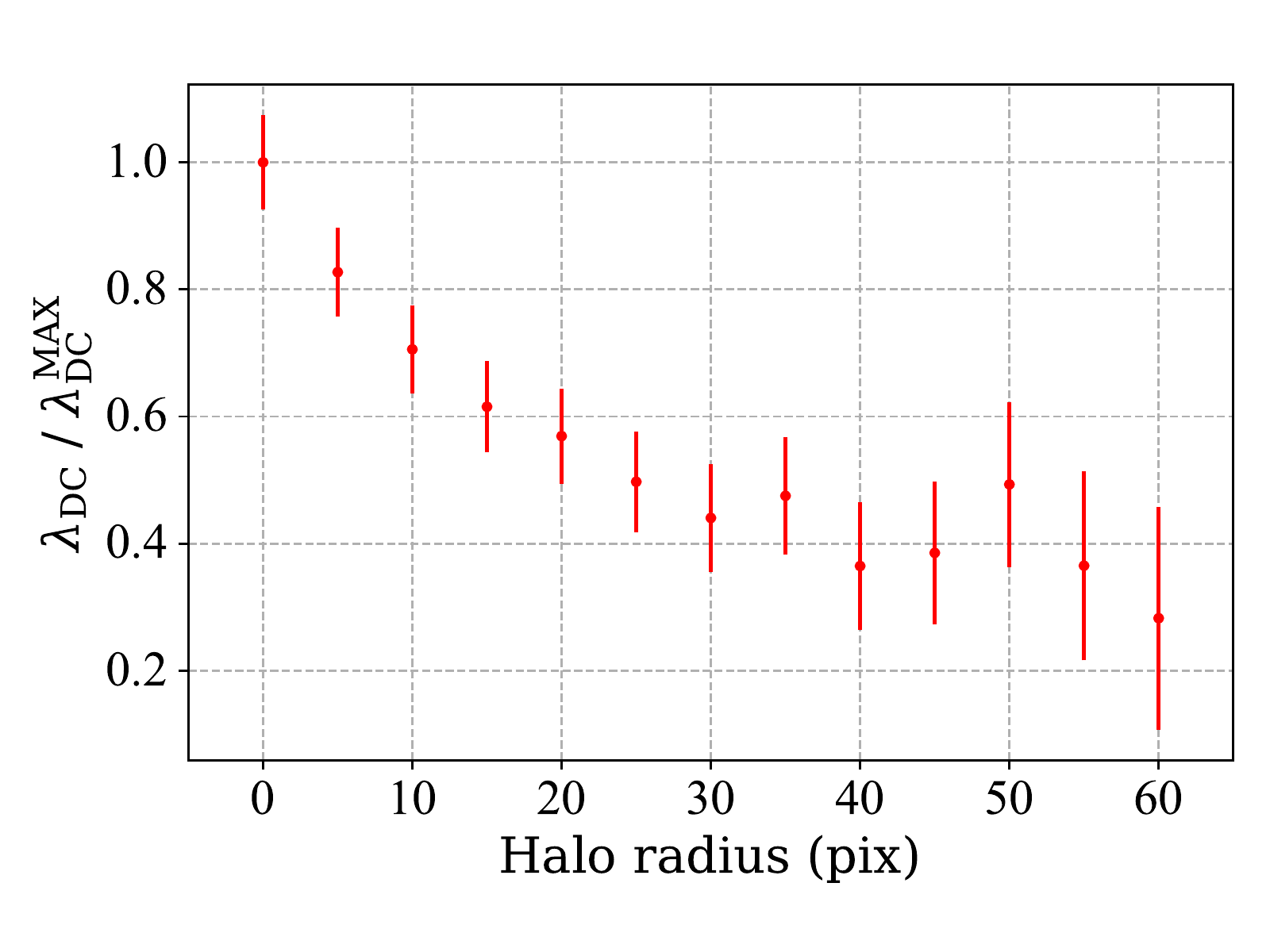}
  \caption{The normalized dark current rate, $\lambda_{\rm DC}$,
  as a function of the halo radius mask. As can be seen, $\lambda_{\rm DC}$ monotonically decreases until a halo radius of roughly 30 pixels.  Dataset~A was used for this figure.}
  \label{fig.Halos}
\end{figure}

As mentioned before, the same event-selection criteria shown in Table~I of~\cite{SENSEI2020} are used, except that the edge and halo mask are set to 40 pixels (instead of 60 pixels) in order to increase statistics (see Fig.~\ref{fig.Halos}). 

\renewcommand{\arraystretch}{1.7}
\begin{table}[t!]
\centering
\caption{Description of each dataset in the section analysis specifying skipper samples (number of samples measured for a given pixel), V$_{\rm DD}$ voltage (voltage applied to the M1 transistor drain channel as shown in Fig.~\ref{fig:RO}), external lead shield presence and both exposure and readout time. These last two, as also the number of skipper samples, vary from image to image in the same dataset as explained in the text.}
\label{table1.3}

\begin{tabular}{lcccc}
\hline
                      & A             & B         & C & D  \\ \hline
Skipper samples       & 250           & 250       & 200-950 & 200-950  \\ 
${\rm V_{DD}}$ voltage (-V)      & 22            & 17-25     & 22      & 21 \\
External shield          & No            & Yes       & No & Yes       \\
Exposure time (hs)    & 0-8 & 0       & 0  & 0   \\ 
Readout time (h:m:s)          & 4:57:00           & 0:55:43       & Variable & Variable      \\ \hline
\end{tabular}
\end{table}

\subsubsection{\label{sec:RDC} Dark current}

\begin{figure}
  \includegraphics[width=1\linewidth]{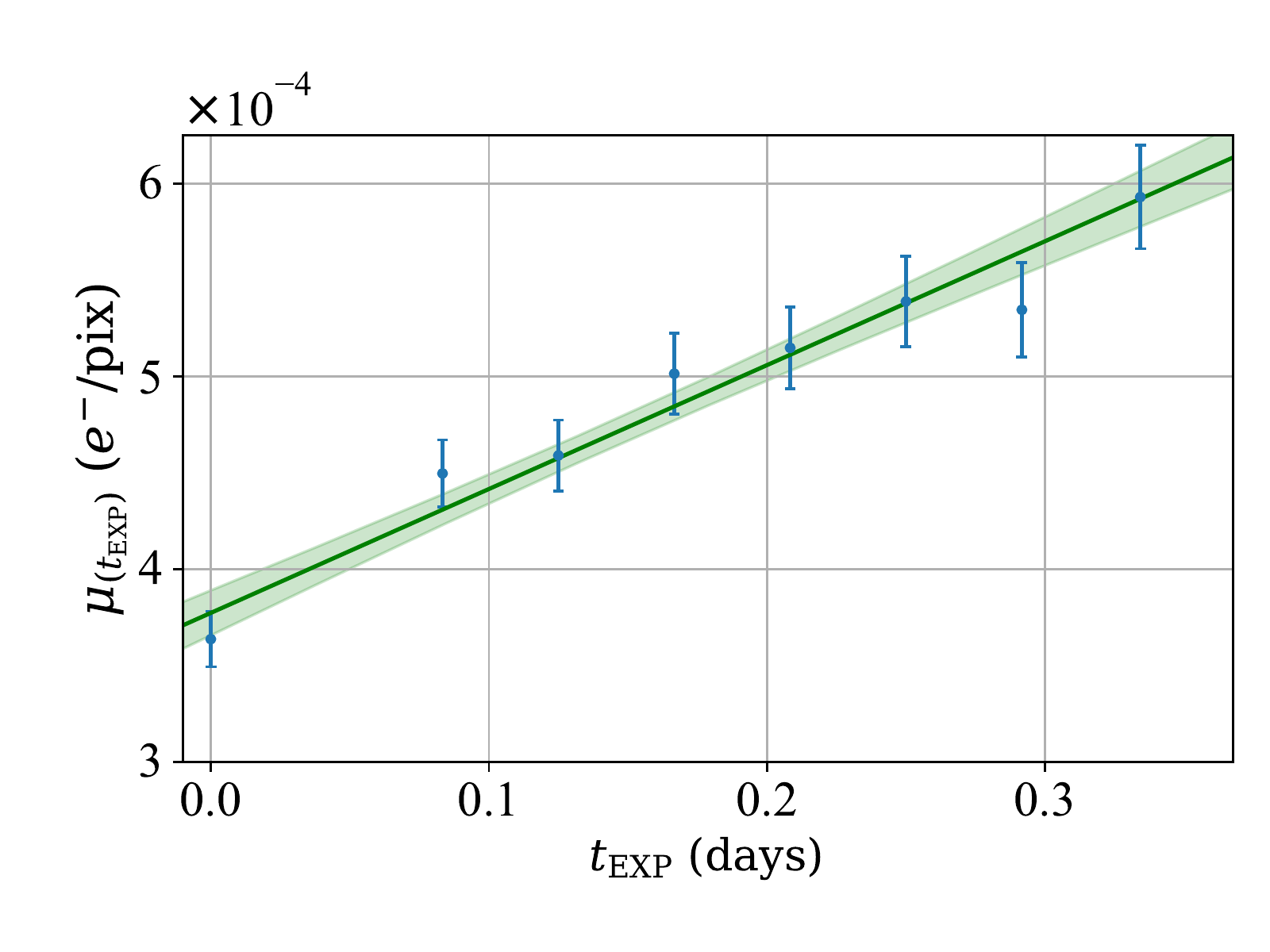}
  \caption{{Determination of $\lambda_{\rm DC}$.} SEE rate as a function of the exposure time $t_{\rm EXP}$ ({\bf blue dots}). The linear regression of these measurements together with a 1$\sigma$ CL band is shown in {\bf green}. The slope is found to be $(5.89 \pm 0.77) \times 10^{-4}$ e$^{-}$ /pix/day and the $y$-intercept  $(3.69\pm0.13) \times 10^{-4} \ e^{-}/{\rm pix} $. Dataset~A was used for this figure.}
  \label{fig.DC}
\end{figure}

Using dataset A, we extract the number of SEE per pixel for each image and perform a linear regression using Eq.~(\ref{eq2.mIII}),  taking the exposure time of each image $t_{\rm EXP}$ as the independent variable. As shown in Fig.~\ref{fig.DC}, we obtain a DC rate of $\lambda_{\rm DC}= (5.89 \pm 0.77) \times 10^{-4} e^{-}/{\rm pix}/{\rm day}$. This value is compatible with the $(5.312^{+1.490}_{-1.277})\times 10^{-4} e^{-}/{\rm pix}/{\rm day}$ reported in~\cite{SENSEI2020} for the configuration without the external shield.

Our measurement identifies a non-negligible value of the time-independent term $(\mu_{\rm RO}+\mu_{\rm SC})$ (see Eq.~(\ref{eq2.mIII})). 
This contribution dominates the number of SEE for exposures shorter than 15~hours. To trace its origin and to disentangle the contributions from $\mu_{\rm RO}$ and $\mu_{\rm SC}$, we perform a dedicated study to understand the SEE produced by luminescence of the output transistor. 
Light generated in the readout stage can reach the pixels in the serial register and active area of the CCD during Readout and produce SEE that are independent of the exposure time. 
The next subsection discusses this characterization effort and how it leads to the optimization of the operation parameters.

\subsubsection{\label{sec:Raievents} Optimization of the operation parameters}

\noindent To characterize the amplifier light contribution, we study the impact on the amplifier light events from varying the voltages in the readout stage. 
We focus on the voltage applied to the V$_{\rm DD}$ gate, which controls the drain voltage of the output transistor M1 (see Fig.~\ref{fig:RO}), as previous works \cite{toriumi1987,lanzoni1991} have shown that an increase in the current between the drain and source leads to an increase in light emission; this phenomenon is explained in the literature as bremsstrahlung radiation produced by hot electrons. These are electrons that, because of very localized high electric fields in the M1 transistor, make a transition from the valence to the conduction band, enabling photon emission as explained in~\cite{toriumi1987}. 

\begin{figure}[t!]
  \includegraphics[width=1\linewidth]{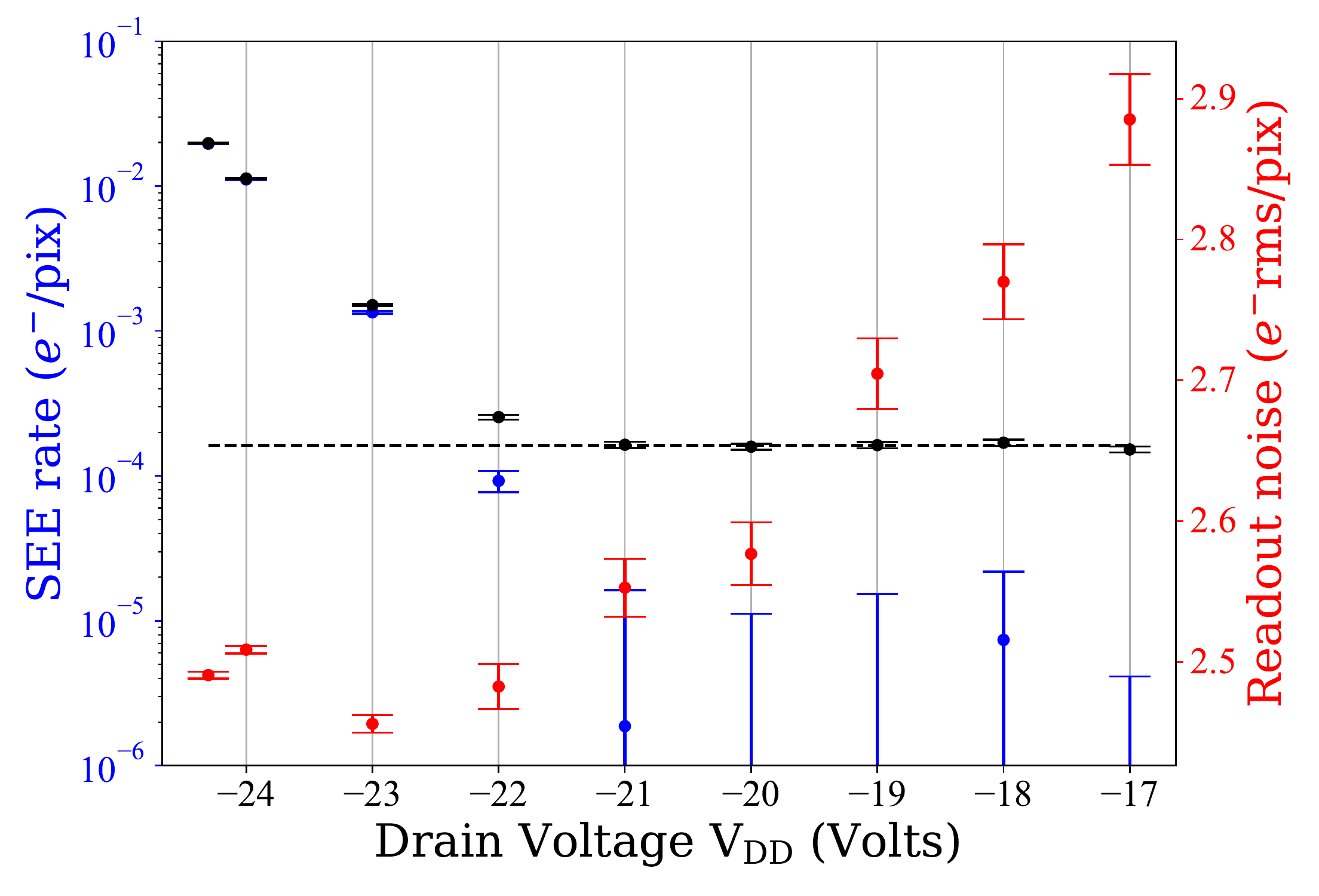}
    \caption{Single-electron events (SEE) per pixel (left axis) and single-sample readout noise (red, right axis) as a function of the drain voltage of the M1 transistor (${\rm V_{\rm DD}}$). In black, we show the SEE per pixel collected for each voltage ($\mu_{(t_{\rm RO})}$) and in blue the AL contribution ($\mu_{\rm AL}$), estimated from Equation \ref{eq2.mIV}. Black dashed line shows estimation for $\mu_{\rm SC}$. Images are taken from dataset B.}
  \label{fig.AIvsVDD}
\end{figure}

To measure the amplifier light contribution we take zero-exposure images for nine different ${\rm V_{\rm DD}}$ values (dataset B) to check if it impacts the number of SEE collected.
As discussed at the end of the previous subsection, light emitted by the output transistor can produce SEE during Readout with a number that is independent of the exposure time. 
By taking zero-exposure images we are able to measure $\mu_{\rm (t_{RO})}$ as a function of ${\rm V_{DD}}$. 
Following Eq.~(\ref{eq2.mIV}), and using the values of $\lambda_{\rm DC}$ and $\mu_{\rm SC}$ from Table \ref{table1.4} (for the corresponding V$_{\rm DD}$ and external shield presence), we can estimate a value of $\lambda_{\rm AL}$ for each ${\rm V_{DD}}$ voltage. The origin of both $\lambda_{\rm DC}$ and $\mu_{\rm SC}$ are explained in the next sub-section.

In Fig.~\ref{fig.AIvsVDD} we show, for each ${\rm V_{\rm DD}}$ voltage in dataset B, the SEE rate of the total events collected ($\mu_{(t_{\rm RO})}$) in black and the AL contribution ($\mu_{\rm AL}$) extracted for each of those voltages in blue, estimated as $\lambda_{\rm AL} \cdot t_{\rm RO}$.
It can be seen that the AL contribution to the overall SEE rate drastically decreases between $-24$~V and $-21$~V while both signals reach a plateau above $-21$~V. This plateau, that corresponds to $\mu_{\rm SC}$ (black dashed line) for $\mu_{(t_{\rm RO})}$, is due to a decrease of light emission in M1 as it is driven from its saturation region (below $-21$~V) to its linear region.
At the same time, a reduction of the V$_{\rm DD}$ value produces an increase on the electronic readout noise, which reduces the signal-to-noise ratio. 
Our technique allows one to optimize the operating conditions depending on the application.

It is not clear due to uncertainties if $\lambda_{\rm AL}$ is zero for values above $-21$~V. To quantify this in a more precise way, we take two sets of images with different $t_{\rm RO}$ (datasets C and D).
These data also allow us to measure the amount of SC events under these conditions.

\renewcommand{\arraystretch}{1.7}
\begin{table}[]
\centering
\caption{Summary of charge contributions and results for ${\rm V_{DD}}$~$=-21$~V and $-22$~V. The units for $\lambda_{\rm DC}$ and 
$\lambda_{\rm AL}$ are $ \ 10^{-4} \ e^{-}/{\rm pix}/{\rm day}$, while $\mu_{\rm SC}$ is in $ \ 10^{-4} \ e^{-}/{\rm pix}$. $\lambda_{\rm AL}$ and $\mu_{\rm SC}$ from the first two rows are extracted from Fig.~\ref{fig.SC} whereas $\lambda_{\rm DC}$ in the first row is obtained from~\cite{SENSEI2020} and in the second row from  dataset~A, as explained above. Additionally, the third row shows $\lambda_{\rm AL}$ extracted from Fig.~\ref{fig.AIvsVDD} as explained in the text.} 

\label{table1.4}
\begin{tabular}{ccccc}
\hline
${\rm V_{DD}}$ & External Shield &  $\lambda_{\rm DC}$ & $\lambda_{\rm AL}$  & $\mu_{\rm SC}$      \\ \hline
$-21$ & Yes &  $(1.59\pm0.16)$  & $(0.36\pm0.18)$  & $(1.52\pm0.07)$ \\
$-22$ & No &  $(5.89\pm0.77)$  & $(19.91\pm1.26)$ & $(1.59\pm0.12)$ \\
$-22$ & Yes &  $-$  & $(23.89\pm3.99)$ & $-$ \\ \hline
\end{tabular}
\end{table}

\subsubsection{\label{section:RSC} Spurious charge and amplifier-light contribution }

\begin{figure}
\centering
  \centering
  \includegraphics[width=1\linewidth]{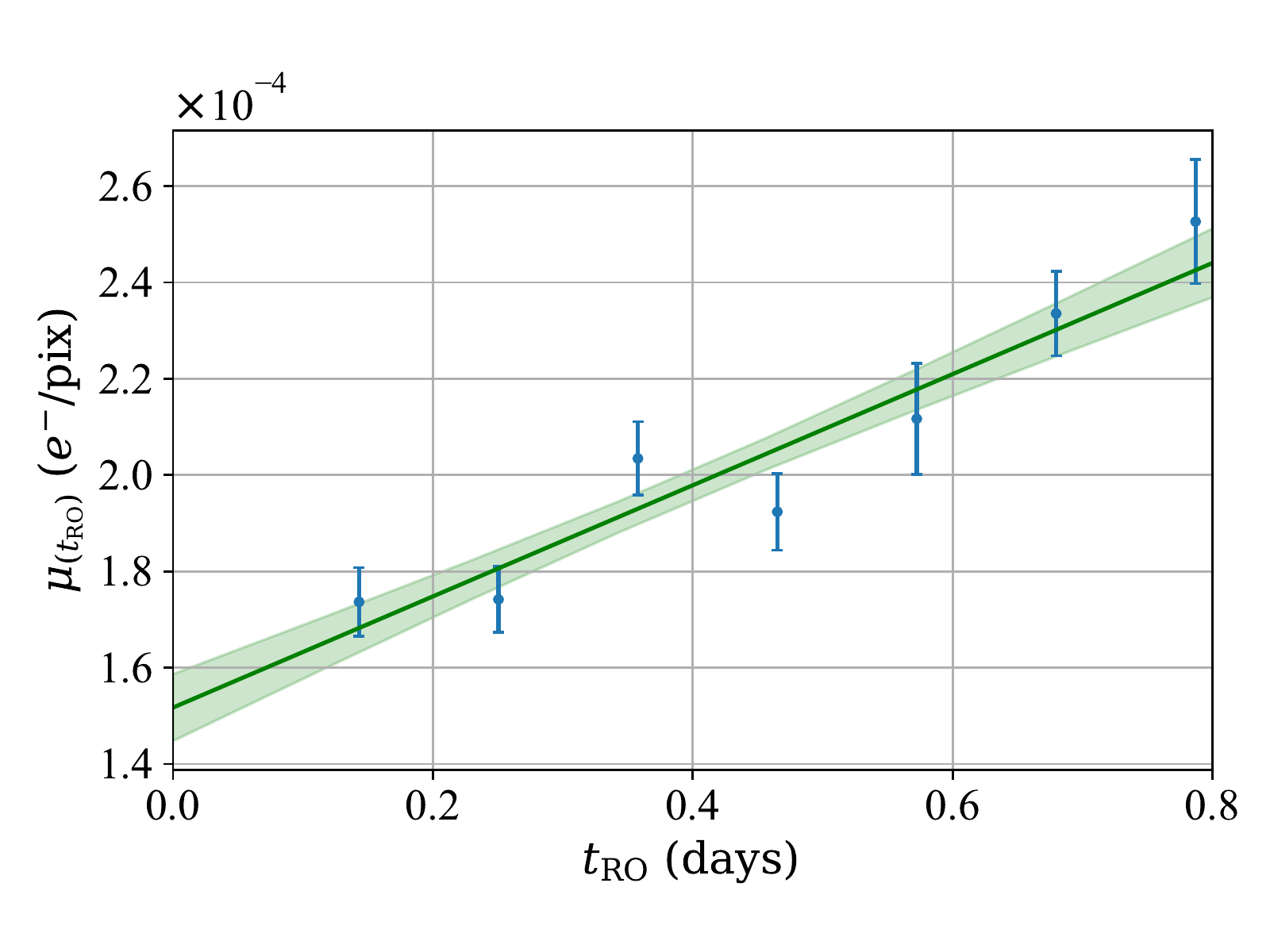}
  \caption{Determination of $\lambda_{\rm AL}$ and $\mu_{\rm SC}$. Extracted SEE rate versus readout time $t_{\rm RO}$ (blue dots) for images in dataset D. 
  In green, the performed linear regression together with a 1$\sigma$ CL band in light green. The extracted value for the slope of the regression is $(1.15 \ \pm \ 0.16) \times 10^{-4} \ e^{-}/{\rm pix}/{\rm day}$ and the y-intercept $(1.52 \pm 0.07) \ \times \ 10^{-4} \ e^{-}/{\rm pix} $.}
  \label{fig.SC}
\end{figure}

To measure the parameters in Eq.~(\ref{eq2.mIV}), the following procedure was applied on both datasets C and D: the amount of SEE per pixel is extracted for each group of images with equal $t_{\rm RO}$ and a linear regression is performed. 
Fig.~\ref{fig.SC} illustrates this result for dataset~D. Following Eq.~(\ref{eq2.mIV}), the slope of the regression is the sum of $\lambda_{\rm DC}/2$ and $\lambda_{\rm AL}$, and the $y$-intercept is the amount of SC ($\mu_{\rm SC}$). 
For $V_{\rm DD} =-22$~V, and using the value for $\lambda_{\rm DC}$ obtained from dataset~A, $\lambda_{\rm AL}$ is determined to be $(19.91\pm1.26) \times 10^{-4} \ e^{-}/{\rm pix}/{\rm day}$. 
For $V_{\rm DD} = -21$~V (the configuration used in~\cite{SENSEI2020}), we use the reported $\lambda_{\rm DC}$ from that work,  $(1.59 \ \pm \ 0.16) \times 10^{-4} \ e^{-}/{\rm pix}/{\rm day}$, as a reference value for $\lambda_{\rm DC}$. 

The first two rows of Table \ref{table1.4} summarize the results after combining Eq.~(\ref{eq2.mIV}) and the values obtained from the linear regression for $V_{\rm DD} =-21$~V and $V_{\rm DD} =-22$~V, as explained in the previous paragraph. 
Additionally, an estimation of $\lambda_{\rm AL}$ extracted from Fig.~\ref{fig.AIvsVDD} is included in the third row in order to compare results for $\lambda_{\rm AL}$ with and without external shield.
As shown, the optimization of the V$_{\rm DD}$ voltage from $-22$~V to $-21$~V reduces the number of amplifier light events by two orders of magnitude. 
%
Additionally, the $y$-intercept values ($\mu_{\rm SC}$) for both datasets C and D (first two rows of Table~\ref{table1.4}) are compatible and do not depend on ${\rm V_{DD}}$, or on the additional shielding, as expected. 
We are currently working on the development of methods to reduce the SC through the shaping of the clock profiles.

\subsection{\label{section:conclusions} Conclusions}

In this work, we provide an empirical model together with the tools to identify and measure the main contributions to single electron events (SEE) in Skipper-CCD sensors. 
Using these techniques, we are able to identify, disentangle, and quantify individual contributions to the SEE rate previously grouped under the label ``dark counts.'' 
Our results determine three types of rates: the dark current (DC), a current of single events produced by amplifier light (AL), and the spurious charge (SC).
In addition, by studying the dependence of the amplifier light on the drain voltage of the output transistor, we reduced this contribution by  two orders of magnitude.
The empirical modeling and techniques presented in this work provide a tool to optimize the operating conditions and push forward the capability of dark matter experiments based on the  Skipper-CCD technology. 

\vspace{5mm}

\begin{acknowledgments}
\noindent\textbf{ACKNOWLEDGMENTS.}
We are grateful for the support of the Heising-Simons Foundation under Grant No.~79921.
This work was supported by Fermilab under DOE Contract No.\ DE-AC02-07CH11359. 
The CCD development work was supported in part by the Director, Office of Science, of the DOE under No.~DE-AC02-05CH11231.
RE acknowledges support from DoE Grant DE-SC0009854, Simons Investigator in Physics Award 623940, and and the US-Israel Binational Science Foundation Grant No.~2016153
The work of DG was supported in part by DoE Grant DE-SC0009854 and BSF grant No.~2016153.  TV is supported by the Israel Science Foundation-NSFC (grant No.~2522/17), by the Binational Science Foundation (grant No.~2016153) and by the European Research Council (ERC) under the EU Horizon 2020 Programme (ERC-CoG-2015 - Proposal n.~682676 LDMThExp).
The work of SU is supported in part by the Zuckerman STEM Leadership Program.
IB is grateful for the support of the Alexander Zaks Scholarship, The Buchmann Scholarship, and the Azrieli Foundation.
This manuscript has been authored by Fermi Research Alliance, LLC under Contract No. DE-AC02-07CH11359 with the U.S.~Department of Energy, Office of Science, Office of High Energy Physics. The United States Government retains and the publisher, by accepting the article for publication, acknowledges that the United States Government retains a non-exclusive, paid-up, irrevocable, world-wide license to publish or reproduce the published form of this manuscript, or allow others to do so, for United States Government purposes. 
\end{acknowledgments}

\begin{appendix}
\section{APPENDIX: Spurious charge in the readout stage.}
\noindent In order to measure the generation of spurious charge (SC) in the readout stage, a dedicated study was carried out using a Skipper-CCD installed at SNOLAB in Sudbury, Canada. This CCD is identical to the one used for the rest of the work (it is part of the same fabrication batch), and the operating conditions (vacuum pressure and CCD temperature) used during the data-taking process were also similar. While the shielding was slightly different, this will not be relevant for our study of SC generation in the readout stage, as will be clear below. Furthermore, the value of VDD during readout was set at $-21$~V.

The main goal of the study was to measure if SC generation in the readout stage is a major contribution to the overall SC. In order to do that, 1146 empty pixels with 5000000 skipper samples each were read clocking only the readout stage. This amounts to $5.73 \times 10^{9}$ samples.

For additional information on the Skipper-CCD operation, we refer the reader to~\cite{janesick} and~\cite{Tiffenberg2017}.

\begin{figure}[h]
\centering
  \centering
  \includegraphics[width=1\linewidth]{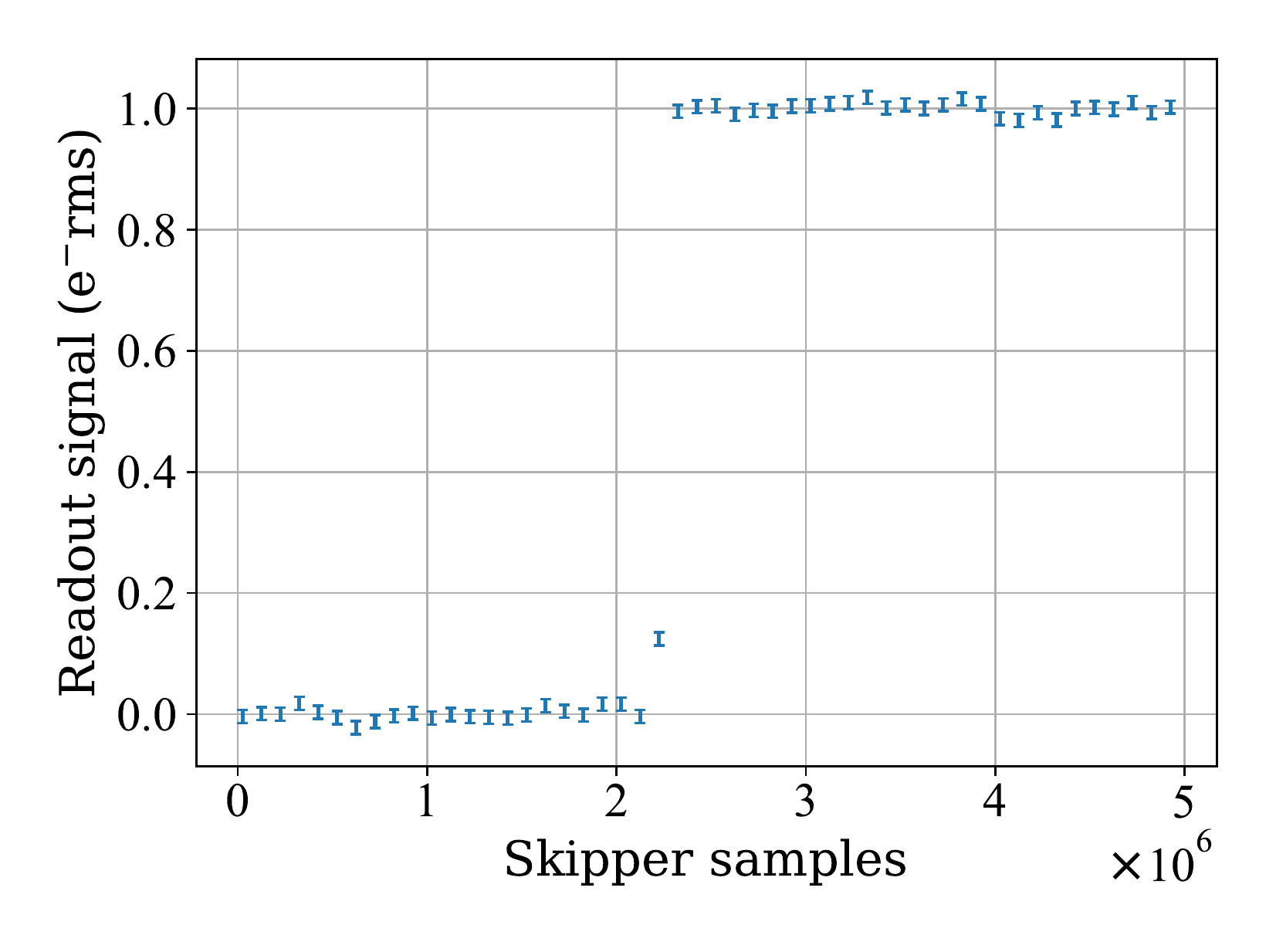}
  \caption{Readout signal for a given pixel versus the number of skipper samples. A ``jump'' in the signal can be seen between 2 and 3 million pixels corresponding to charge generation.} 
  \label{fig.SN}
\end{figure}

To identify the generation of charges during this process, all pixels were analyzed to look for "jumps" in the signal, i.e., charge generation. This was done by dividing the 5000000 samples of each pixel into 50 groups of 100000. A "jump" was identified if the mean of a group was away from the next one by at least 5 standard deviations of the former group but less than 100e-, an energy threshold from which it was considered that a high energy event had interacted with the readout stage during the skipper process. An example of a jump is shown in Fig.~\ref{fig.SN}.

Within the $5.73 \times 10^{9}$ samples dataset, 51 ``jumps'' were found, giving us an event rate of $(8.9\pm 1.3) \times 10^{-9} \ {\rm events / \rm sample}$ if we attribute all these events to SC generated in the readout stage.

For the data collected in the main part of the paper, the number of samples were always less than 1200 samples.  This would produce a SC rate of $(1.1\pm0.2) \times 10^{-5} \ {\rm events / \rm pix}$ or ${\rm e^{-} / \rm pix}$, which is negligible when compared to overall SC contribution. For a regular DM science-run, for which $\sim 300-400$ skipper samples are taken~\cite{SENSEI2020}, a SC rate of $\sim 3 \times 10^{-6} \ {\rm e^{-} / \rm pix}$ is expected from the readout stage.

It is yet to be understood how many of these events come from luminescence from the readout stage itself and how many from actual SC.

\end{appendix}

\bibliography{bibliography.bib}

\end{document}